\documentclass[a4paper, amsfonts, amssymb, amsmath, reprint, showkeys, nofootinbib, twoside]{revtex4-2}
\usepackage{lipsum}
\usepackage{mathtools}
\usepackage{graphicx}
\usepackage{dcolumn}
\usepackage{amsmath}    
\usepackage{amssymb}
\usepackage{bm}
\usepackage{hyperref}
\usepackage{latexsym}
\usepackage{verbatim}
\usepackage{color}
\usepackage{float}
\usepackage[caption=false]{subfig}
\usepackage{epstopdf}

\newcommand{\ndg}[1]{\textcolor{black}{#1}}
\newcommand{\sk}[1]{\textcolor{black}{#1}}
\def\beq{\begin{equation}}
\def\eeq{\end{equation}}

\def\beq{\begin{equation}}                          
\def\eeq{\end{equation}}                          
\def\bea{\begin{eqnarray}}                          
\def\eea{\end{eqnarray}}

\DeclareRobustCommand{\uvec}[1]{{%
  \ifcsname uvec#1\endcsname
     \csname uvec#1\endcsname
   \else
    \bm{\hat{\mathbf{#1}}}%
   \fi
}}
\usepackage[english]{babel}
                   
\preprint{}

\begin{document}

\title{Hydrodynamic bend instability of motile particles on a substrate}


\author{Sameer Kumar$^\dag$}
\affiliation{Niels Bohr Institute, University of Copenhagen, Copenhagen, Denmark}
\affiliation{Department of Physics, Indian Institute of Technology Kanpur, Kanpur, India}

\author{Niels de Graaf Sousa$^\dag$}
\affiliation{Niels Bohr Institute, University of Copenhagen, Copenhagen, Denmark}

\author{Amin Doostmohammadi}
\email[]{doostmohammadi@nbi.ku.dk}
\affiliation{Niels Bohr Institute, University of Copenhagen, Copenhagen, Denmark}

\begin{abstract}
The emergence of hydrodynamic bend instabilities in ordered suspensions of active particles is widely observed across diverse living and synthetic systems, and is considered to be governed by dipolar active stresses generated by the self-propelled particles. Here, using linear stability analyses and numerical simulations, we show that a hydrodynamic bend instability can emerge in the absence of any dipolar active stress and solely due to the self-propulsion force acting on polar active units  
suspended in an incompressible fluid confined to a substrate. 
Specifically, we show analytically, and confirm in simulations, that a uniformly ordered state develops bend instability above a critical self-propulsion force. Numerical simulations show that a further increase in the self-propulsion strength leads the system towards a disorderly flow state. 
The results offer a new route for development of hydrodynamic instabilities in two-dimensional self-propelled materials that are in contact with a substrate, with wide implications in layers of orientationally ordered cells and synthetic active particles.
\end{abstract}
\maketitle

\def\thefootnote{\dag}\footnotetext{These authors contributed equally to this work.}\def\thefootnote

\section{Introduction}
Collective motion is ubiquitously observed in active materials such as schools of fish \cite{VicsekPRL1995}, birds flock \cite{TonerPRL1995}, bacterial colony \cite{DombrowskiPRL2004, SokolovPRL2007, PeruaniPRL2012, WiolandPRL2013} and in the coordinated movement of cells in tissues \cite{SerraPicamal2012, KemkemerEPJE2000, SawNature2007, HeinrichElife2020}. The emergence of collective behavior in such living systems has further inspired realization of synthetic and reconstituted active materials that are characterized by collective coordinated motion of their constituents
\cite{KudrolliPRL2008, DeseignePRL2010,SchallerNature2010, SuminoNature2012, SanchezNature2012,GuillamatPNAS2016,BechingerRMP2016}. In this vein, not only understanding collective migration is central to various biological functions that are associated with coordinated motion of cells in tissues \cite{WangIOVS2003, FriedlJCB2009, HidalgoNatCB2011, FriedlNatCB2014, MishraDevelopement2019,HeinrichElife2020}  and bacteria in biofilms \cite{ZhangPNAS2010, IshikawaAPLBio2020, LiNatCom2021}, it provides a potential for design and control of materials that are capable of self-organization and self-propulsion \cite{LiebchenACR2018, JiAdvMat2023}.
These widespread implications of collective motion in living, synthetic, and bioinspired materials has in turn triggered development of diverse theoretical models to provide a fundamental understanding of the mechanisms that control various modes of collective pattern formation \cite{MarchettiRMP2013, BechingerRMP2016, JülicherRPP2018}.

While particle based models have been pivotal in understanding various forms of collective motion,~\cite{VicsekPRL1995, GregoirePRL2004, BuhlScience2006}, coarse-grained, continuum descriptions have been instrumental in unveiling generic mechanisms that govern the emergence of collective patterns of motion~\cite{VicsekPR2012, RamaswamySimhaToner2003, TonerPRL1995, TonerAoP2005}. In particular, coupling the orientation field associated to the motion or alignment of the active particles to their self-generated flow fields has been successful in describing various degrees of complexity in the emergence of collective motion, from fundamental instabilities that initiate the collective flows \cite{TonerPRL1995, SimhaRamaswamyPRL2002}, to statistical properties of the active flows \cite{Li2019, MartinezPRX2021}. In these descriptions the orientational order typically accounts for two distinct symmetries: ({\it i}) the polar symmetry that can be associated with either the directional motion of the particles or their inherent head-to-tail asymmetry \cite{BrotoPRL2013,BartoloPRX2021}, such as planar cell polarity in eukaryotic cells \cite{FisherDevelopement2019} or asymmetry of Janus particles \cite{YanMatToday2021}, and ({\it ii}) apolar, nematic symmetry that is associated with the alignment of elongated active particles, such as self-propelled rods \cite{KudrolliPRL2008}, bacteria~\cite{volfson2008}, or direction of the cell shape elongation in epithelial cells~\cite{SawNature2007}.
%

In a dense active system, coupling the orientation field to the flows generated by self-propelled particles has revealed that the emergent collective flows are governed by a generic hydrodynamic instability, inducing orientational deformation modes, that are dependent on the nature of the active stress that particles generate~\cite{SimhaRamaswamyPRL2002, RamaswamyNJP2007, EdwardsEPL2009}.  This active stress, based on the symmetry arguments, is dipolar in nature \cite{PedleyARF1992}, and in two dimensions induces bend (or splay) instabilities for extensile (contractile) particles which push (pull) flows along their elongation axis \cite{SimhaRamaswamyPRL2002}. 

The discovery of hydrodynamic bend-splay instabilities in two-dimensional suspensions of orientationally ordered active particles marks one of the milestones in understanding collective behavior in active materials~\cite{SimhaRamaswamyPRL2002,MartinezNatPhys2019}. Two of the defining features of this well-established hydrodynamic instability are: (i) it is for an unbound system, and (ii) the instability is governed by dipolar active stresses. 
In this regard it is important to note that in many experimental realizations of active matter in two-dimension, living or synthetic, active particles interact with a substrate and dissipate a part of energy to the substrate friction \cite{Parrish1997, SerraPicamal2012, KudrolliPRL2008, SurreyScience2001}. Furthermore, while many living and synthetic active particles migrate across substrates due to self-propulsion, models that include only dipolar active stresses, without explicit self-propulsion forces, capture the case of active particles which generate flows through force dipoles but do not migrate. \ndg{Such models have been instrumental in showing the role of active stresses in driving instabilities and pattern formation, particularly in systems with nematic order. However, in systems where persistent motility and directed transport are central to the dynamics, neglecting self-propulsion can overlook important aspects of force transmission and large-scale organization.} Previous studies of \em{dilute} suspensions of \em{compressible} active particles in confined thin films have shown the suppression of bend instabilities and instead predict unstable splay modes that depend on the head-tail shape asymmetry of the active particles~\cite{BrotoPRL2013, LefauvePRE2014}. 
Indeed, a recent theoretical study of \sk{\em{incompressible}} active polar system in contact with substrate has shown that in the absence of dipolar active stress, the bend-splay instabilities of the ordered state are suppressed and instead the system exhibits anomalous stability and long-range order \cite{MaitraPRL2020}. \sk{Moreover, a model of expanding epithelial cells on a substrate as a compressible self-propelled polar matter is shown to undergo an instability from an isotropic state to spontaneously flowing state \cite{HeinrichElife2020}. Including dipolar stress in such a model for epithelial cell sheets is shown to result in non-linear wave-propagation and turbulent dynamics~\cite{blanch2017hydrodynamic}.}
\begin{figure}[t!]      
\includegraphics[width=\linewidth]{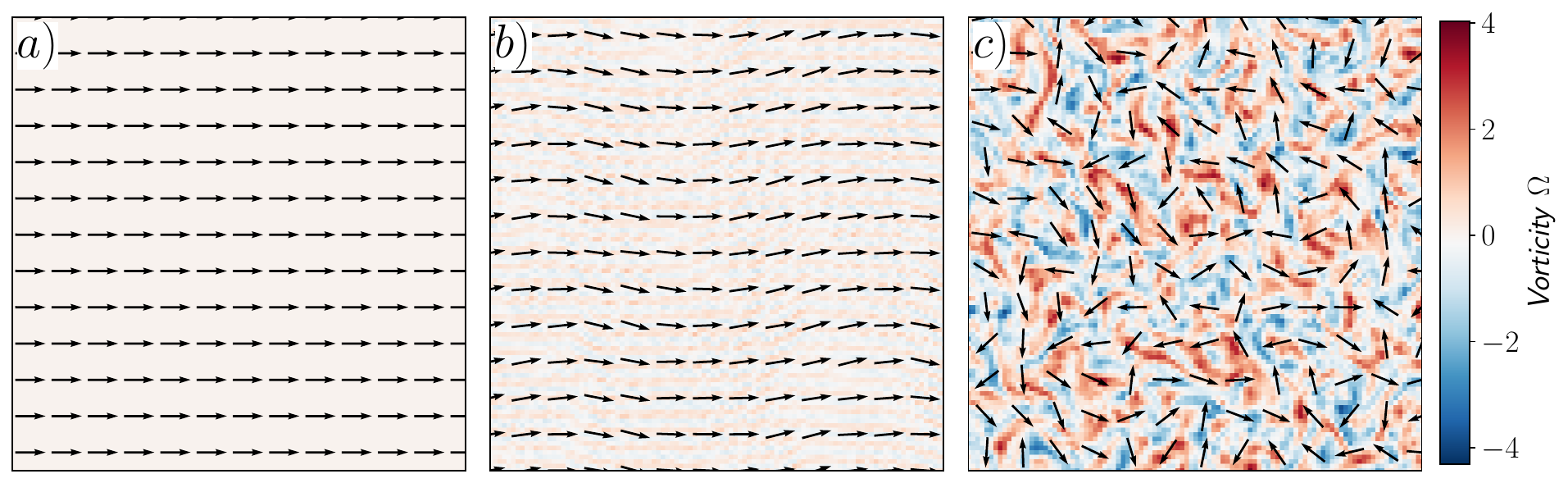}
\caption{\ndg{\textbf{Simulation snapshots of the three distinct states} obtained from the simulations of the full model. (a) Flowing state for $\alpha=0$ and $\beta=0.5$, (b) undulating or bending state for $\alpha=2.5$ and $\beta=0.8$, and (c) disorderly flow for $\alpha=4.0$ and $\beta=0.2$. Colormap denotes the systems vorticity $\Omega$. Only small portion of the full simulation domain is shown for clarity.}} 
\label{fig: snapshots}
\end{figure}

In this letter, we present a new route for the development of hydrodynamic bend instabilities in ordered suspensions of \sk{incompressible} active particles in contact with a substrate and {\it in the absence of} dipolar active stress. Based on the numerical simulation and the linear stability analysis, we find that the self-propulsion force, \sk{as the only source of activity,} can lead to a bend instability in an ordered suspension of active polar particles in contact with a substrate, and show that there exists a critical self-propulsion threshold that delineates the boundary between the stable and unstable variable phase. We further represent a stability diagram of the system and unveil the underlying mechanism of this instability.

{\em Model:-}
\label{model}

We use general dynamical equations for the polarization field, ${\bm p}({\bm r},t)$, of active units suspended
in a fluid with the total velocity field of the particles and
the fluid being ${\bm v}({\bm r},t)$, where ${\bm r}$ is a two-dimensional position vector. The joint density, $\rho$ of the particles and
the fluid is conserved, i.e., $\dot{\rho}= 0$ implying $\partial_i v_i=0$.
In the absence of activity and fluid flow, the equilibrium relaxation derives from a Landau–de Gennes free energy given as,
$\mathcal{F}=\int dA[C(1-\lvert p \rvert^2)^2+\frac{K}{2} \vert \nabla {\bm p} \vert^2]$, where $C$ sets the energy scale for Landau-de Gennes free energy \cite{PGdeGenneBook1995}and $K$ is the Frank elastic constant within the single elastic constant approximation~\cite{FrankRSC1958}. The equation for polarization, ${\bm p}$ is, 

\begin{equation}
    \partial_t {p}_i+ v_k \partial_k p_i+\Omega_{ij}p_j =\frac{1}{\gamma}{h}_i + \beta v_i+\lambda E_{ij}p_j,
\label{eq: minimal equation polarity}
\end{equation}
where $\gamma$ is the rotational viscosity that controls relaxation of the polarity field to the minimum of the free energy through the molecular field $h_i=-\delta \mathcal{F}/\delta{p_i}$. $\Omega_{ij}=\frac{1}{2}(\partial_i v_j-\partial_j v_i) $ and $ E_{ij}=\frac{1}{2}(\partial_i v_j+\partial_j v_i)$ are the vorticity and strain rate tensors. The terms on the lhs define the comoving and co-rotating derivatives. The coefficient $\beta$ in Eq.~(\ref{eq: minimal equation polarity}), aligns the polarization vector with the local suspension velocity \cite{MaitraPRL2020} and is specific to systems in contact with a substrate \cite{ BrotoPRL2013,NKumarNatCom2014,tsang2016density}. \ndg{The $\beta$ term, models the momentum transfer between the fluid flow and the particles, this generates a torque that tends to align the particles with the flow~\cite{BrotoPRL2013,NKumarNatCom2014}.}
\begin{figure}[b!] 
\includegraphics[width=\linewidth]{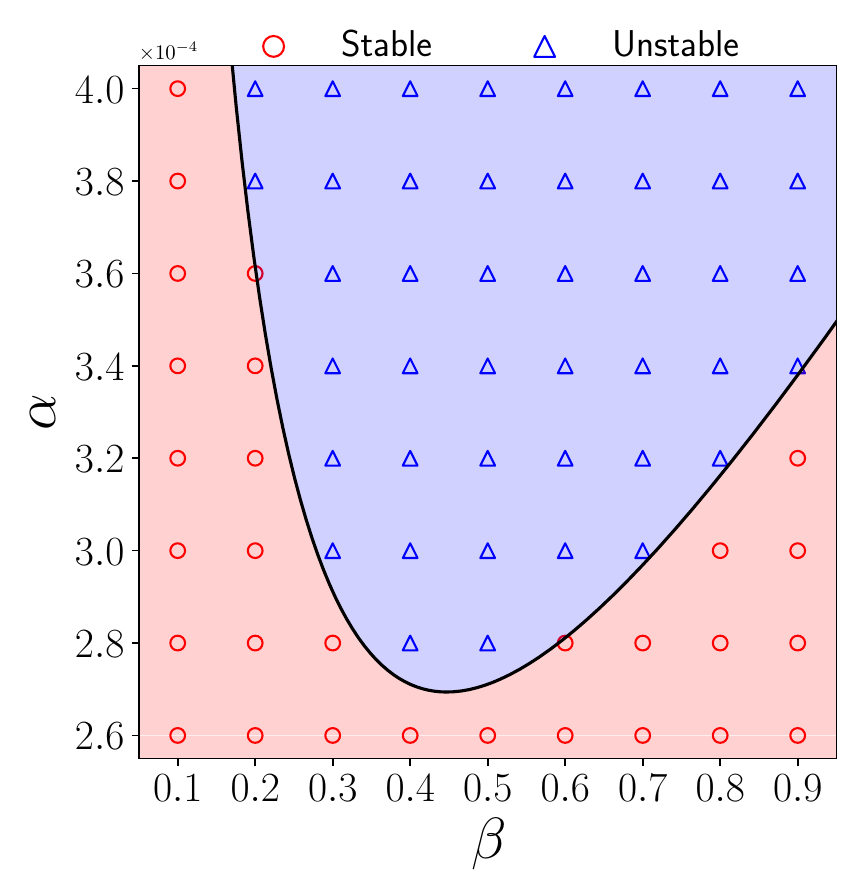}
\caption{\ndg{ \textbf{Stability diagram of the $\mathbf{(\beta,\alpha)}$ phase space}. The blue triangles and red circles denote the unstable and stable states, respectively, determined by simulations. The theoretical prediction of the linear stability analysis, $\alpha_c$ (Eq.~\ref{eq: critical self-propulsion}), is shown as a black solid line, with $q=\frac{2 \pi}{L}$.}}
\label{fig: Stability Diagram}
\end{figure}

The equation governing the dynamics of the velocity field is given by the Navier-Stokes equation,
\begin{equation}
\rho(\partial_t + v_k\partial_k)v_i  =  \alpha p_i -f v_i +\beta h_i+\partial_j \sigma_{ij},
\label{eq: minimal NS eq}
\end{equation}
where, $f$ is the friction coefficient. $\alpha {\bm p}$ is the only active term characterizing the self propulsion in the system that describes the migration of active particles along their polarity direction ${\bm p}$, and the coefficient $\alpha$ sets the self-propulsion strength. In the absence of any active force, the term $\beta {\bm h}$ is required to fulfill the detailed balance condition compensating for the $\beta {\bm v}$ term in Eq.~(\ref{eq: minimal equation polarity}) \cite{MaitraPRL2020}. The stress tensor, $\sigma_{ij}$, encompasses both the viscous stress, $\sigma_{i j}^{\text{viscous}}=2 \eta E_{i j}$ with $\eta$ being the viscosity coefficient, and the elastic stress $\sigma_{i j}^{\text{elastic}}=-P \delta_{i j}+\frac{1-\lambda}{2}p_i h_j - \frac{1+\lambda}{2}p_j h_i + \frac{\lambda}{2} p_k h_k \delta_{i j}$, with $P$ being the isotropic pressure~\cite{GiomiSM2012}.
\sk{We omit the coupling to the concentration field, as it does not qualitatively alter the characteristic states observed, nor does it affect the instability, see Appendix \ref{ConcEffect}.}

{\em Results:-}
\label{results}
We first present results from the simulation, where Eq.~(\ref{eq: minimal equation polarity}) and Eq.~(\ref{eq: minimal NS eq}) are numerically solved using \ndg{OpenFOAM-v2306}. The simulations are performed in a square lattice of length $L$ with Periodic Boundary Conditions (PBC). The values used throughout this study are the following: $\gamma=10.0$, $f=2.0$, $\eta=1.0$, $\lambda=0.3$, $C=0.1$, $K=0.5$, $\rho=1.0$ and the system size is set to $L=256$. \ndg{The parameters are chosen to ensure that the Reynold number $Re \ll 1$.} 
\begin{figure}[b!]      
\includegraphics[width=0.5\textwidth]{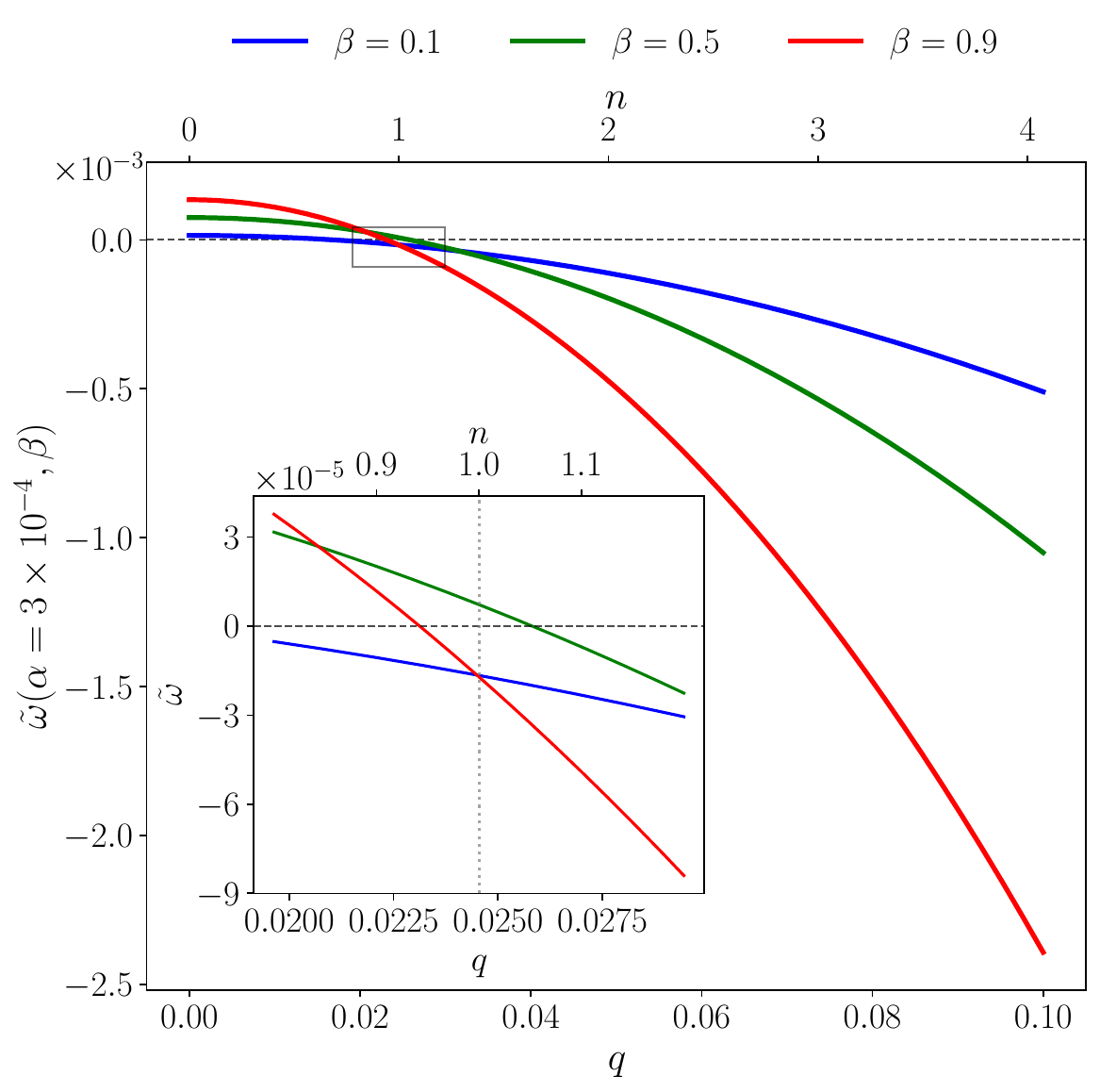}
\caption{\ndg{\textbf{Dispersion relation}. Growth rate $\tilde{\omega}$ as a function of the wave-vector $q$ and wavenumber $n$ for three alignment coefficients ($\beta = 0.1, 0.5, 0.9$) at fixed self-propulsion strength $\alpha = 3 \times 10^{-4}$. The inset shows an enlarged view of the stability-determining region for the fastest-growing mode.}} 
\label{fig: Dispersion Relation Study}
\end{figure}
Starting from a flowing $x$-aligned initial configuration, we simulate the system for different self-propulsion strength, $\alpha$ and velocity coupling coefficient $\beta$. For a chosen set of the self propulsion strength, $\alpha$ and the velocity coupling coefficient, $\beta$, the system leads to the emergence of three qualitatively distinct states: the (i) {\em flowing state} where the system remains in the aligned state (Fig.~\ref{fig: snapshots}a), the (ii) {\em undulating state}  (Fig.~\ref{fig: snapshots}b) and the (iii) {\em disorderly flows state} (Fig.~\ref{fig: snapshots}c). In the flowing state, the polar units remains completely aligned in the steady state. This state is reminiscent of the flocking phase widely observed in Vicsek-type interaction mechanisms with low noise \cite{VicsekPRL1995, TonerPRL1995}, as well as the anomalous stable state reported by Maitra et al~\cite{MaitraPRL2020}. As the strength of self-propulsion is increased beyond a certain threshold, the polar units start undulating in a coherent manner which results in a bend pattern that we refer to as  the {\em undulating state}. 
In the undulating state, the polarity of the active units exhibit fluctuations perpendicular to the initially ordered configuration ($\hat{\bm y}$-direction in Fig.~\ref{fig: snapshots}(b)), resulting in the formation of travelling waves of bend deformation along the ordered direction ($\hat{\bm x}$-direction in Fig.~\ref{fig: snapshots}(b)). While, at the first sight, this undulating state might show similarities to the ``traveling wave" state reported for active polar fluid~\cite{GiomiSM2012}, it is important to emphasize that the bend instability leading to undulations is of a fundamentally different nature since the traveling waves reported in \cite{GiomiSM2012} were the results of the well-established hydrodynamic instability caused by dipolar active stress of the form $\sim (p_i p_j - \frac{p^2}{2} \delta_{ij})$ and their crosstalk by the self-advection in the polarity field. A further increase in the self-propulsion strength leads the system towards the  {\em disorderly flow} state, where the deformations of the polarity field evolve to span all directions (Fig.~\ref{fig: snapshots}(c)).

{\em Linear stability analysis:-}
\label{stability} 
To rigorously characterize the observed hydrodynamic bend instability driven solely by self-propulsion force, we examine the stability of the minimal model that couples the following dynamic variables: the two components of the polarization, $p_x$ and $p_y$, and the vorticity, $\Omega$. The vorticity is defined as $\Omega=\partial_x v_y-\partial_y v_x$ and alongside the steam function $\psi$ given by $v_x=\partial_y \psi$ and $v_y=-\partial_x \psi$ ensures the incompressibility condition. \ndg{To isolate the essential mechanism underlying the instability, we neglect co-rotational terms that couple polarity dynamics to velocity gradients, as well as advective terms in the polarity and velocity equations. This simplification allows us to focus on the interplay between self-propulsion and flow alignment with the minimal ingredients required to trigger the instability. In the Appendix \ref{stability1}, we show that including the full terms does not suppress the instability, underscoring the robustness of the underlying mechanism.} Under these conditions, one can then write the Navier-Stokes equation (Eq.~\ref{eq: minimal NS eq}) in terms of vorticity
\begin{equation}
    \label{eq: NS in terms of vorticity}
    \rho \partial_t \Omega= \alpha (\partial_x p_y - \partial_y p_x)+ \beta (\partial_x h_y - \partial_y h_x) -f \Omega.
\end{equation}

With the dynamic variables formally defined, we examine the linear stability of a uniformly aligned initial configuration, $\phi^{0}=(p_{x}^0, p_{y}^0, \Omega^{0})=(1,0,0)$ \ndg{and apply an} infinitesimal perturbation $\delta \phi^{(1)}({\bf r},t)=(\delta p, \theta, \Omega)$ such that the total system takes the following form ${\bf \phi(r},t) = \phi^0 + \delta {\bf \phi^{(1)}(r},t)$. The perturbation applied to the system reads
\begin{equation}
{\phi^{(1)}(\bf r},t) =   \phi_{nm}(t)e^{i (q_nx+q_my)+\tilde{\omega}t},
\label{eq: Perturbation equation}
\end{equation} 
which accounts for the PBC and includes the growth rate of the perturbation $\tilde{\omega}$. The wave vectors are given by $q_n=\frac{2\pi}{L}n$ and $q_m=\frac{2\pi }{L}m$, with $q^2=q_n^2+q_m^2$. Introducing the perturbation (Eq.~\ref{eq: Perturbation equation}) into the dynamical equations (Eqs.~\ref{eq: minimal equation polarity} and \ref{eq: NS in terms of vorticity}) yields the following system of coupled equations: 

\begin{equation}
    \label{eq: px pert}
    \delta p \left(\frac{Kq^2}{\gamma}+ \tilde{\omega}\right)=i \frac{q_m}{q^2}\beta \Omega,
\end{equation}
\begin{equation}
    \label{eq: py pert}
     \theta \left(\frac{Kq^2}{\gamma}+ \tilde{\omega}\right)=-i \frac{q_n}{q^2}\beta \Omega,
\end{equation}
\begin{equation}
    \label{eq: vortc pert}
    \Omega(\rho \tilde{\omega}+f)=i( \alpha-\beta Kq^2)[q_n \theta -q_m \delta p].
\end{equation}

\ndg{Note that we exploit the coupling between vorticity and the stream function via the Poisson relation, $\Delta \psi=-\Omega$, which implies a direct relationship between their Fourier components, $\psi_{nm}=q^{-2}\Omega_{nm}$. The perturbation amplitudes naturally drop out of the system of equations (Eqs.~\ref{eq: px pert}, \ref{eq: py pert} and \ref{eq: vortc pert}), yielding the following dispersion relation}

\begin{equation}
    \label{eq: dispersion relation}
    \ndg{\rho \tilde{\omega}^2+ \left[f+ \rho \frac{K}{\gamma}q^2\right] \tilde{\omega}+\frac{K}{\gamma}f q^2=\beta(\alpha-\beta K q^2).}
\end{equation}

This relation describes the evolution of the perturbation through a non-linear relation between the wave-number $q$ and the growth rate $\tilde{\omega}$ ( See Appendix \ref{FullOmega} for the full expression). By differentiating Eq.~(\ref{eq: dispersion relation}) with respect to the wavenumber, we find that the growth rate reaches its maximum at the smallest wavenumber, $q=0$, similar to the hydrodynamic instabilities generated by dipolar flows in both polar \cite{GiomiSM2012} and nematic \cite{MuhuriEPL2007,GiomiIOP2012} active systems. The fastest-growing mode corresponds to the wavenumber $q=\frac{2\pi}{L}$,  specifically the longitudinal mode, with $n=1$ and $m=0$. In Fig.~\ref{fig: Dispersion Relation Study} we display the growth rate curves, for a constant self-propulsion coefficient and varying $\beta$. The critical self-propulsion $\alpha_c$ that separates the stable ($\tilde{\omega}<0$) from the unstable ($\tilde{\omega}>0$) states can be found by examining when $\tilde{\omega}$ crosses zero. This leads to the following relation
\begin{equation}
    \label{eq: critical self-propulsion}
    \alpha_c=Kq^2 \left(\frac{f}{\gamma}\frac{1}{\beta}+\beta\right),
\end{equation}
which is displayed as a black curve in Fig.~\ref{fig: Stability Diagram} and reveals close agreement with the simulation results. From this expression, we identify the characteristic coupling coefficient $\beta^* = \sqrt{\frac{f}{\gamma}}$ that delineates two distinct regimes: (i) for $ \beta < \beta^* $, the critical self-propulsion shows asymptotic dependence with $\beta $, while (ii) for $ \beta > \beta^* $, it varies linearly with $\beta $. The inset in Fig.~\ref{fig: Dispersion Relation Study} shows an enlarged view of the first growing mode region centered at $q=\frac{2 \pi}{L}$. For alignment coefficients $\beta = 0.1$ and $0.9$ (\textit{blue} and \textit{red} curves, respectively) the growth rate at the first mode ($n=1$) is negative, indicating decaying perturbations and system stability. In contrast, the case with $\beta=0.5$ (\textit{green} curve) exhibits a positive growth rate, leading to growing perturbations and system instability.

\begin{figure}[b!]
\includegraphics[width=\linewidth]{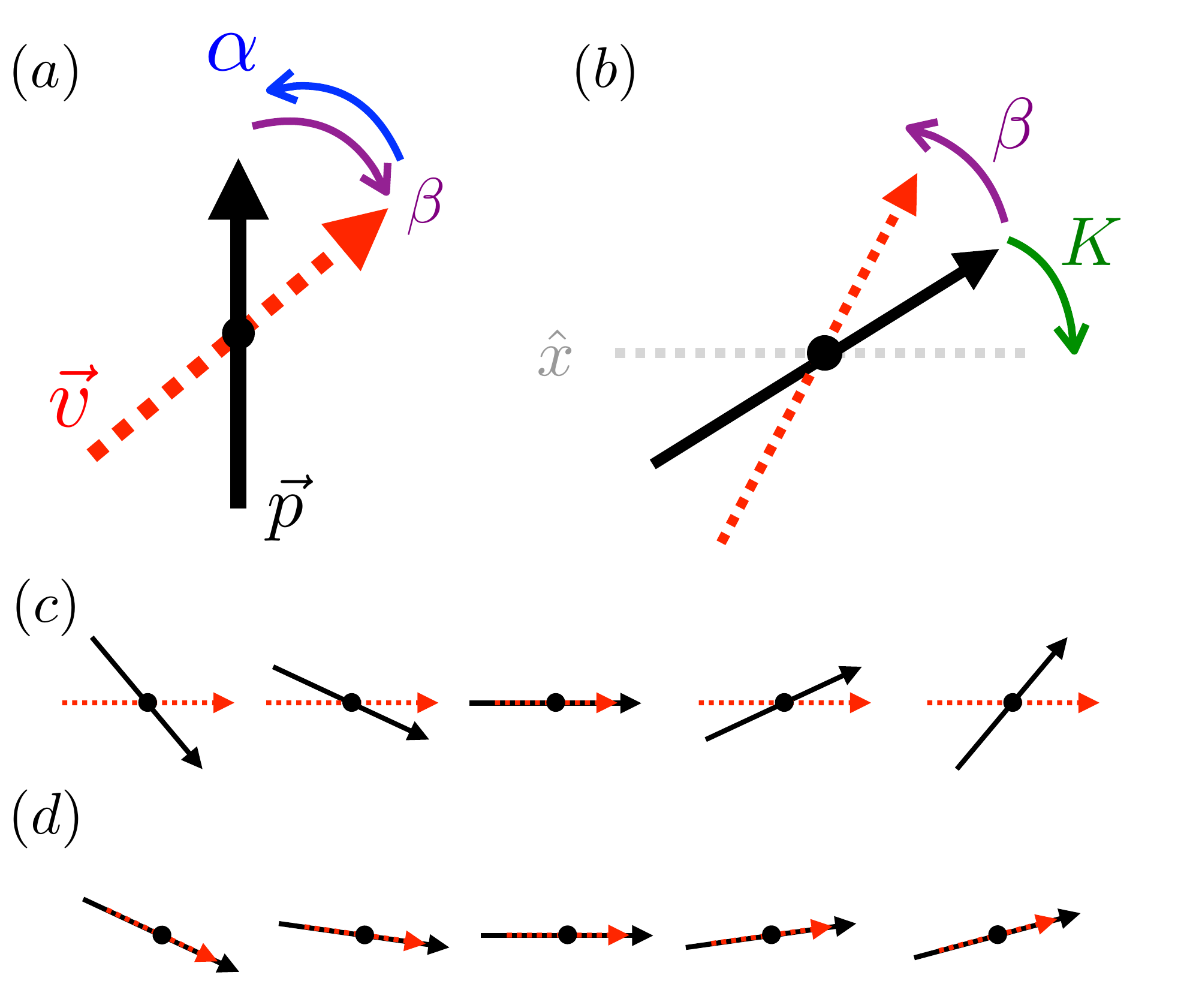}
\caption{ \ndg{\textbf{Schematic of the instability mechanism.} (a) Contributions of the self-propulsion, $\alpha$ and the flow aligning term $\beta$. (b) Schematic representation of the competition between the Frank elastic energy and the flow aligning term. (c) Initial conformation and (d) conformation after the aligning contributions. Black arrows indicate the polar vector $\vec{p}$ an red dashed arrows indicate the flow velocity vector $\vec{v}$}.} 
\label{fig: Mechanism Schematic}
\end{figure}

{\em Mechanism:-} The relation that determines the systems stability, given by Eq.~(\ref{eq: critical self-propulsion}), reveals the three key components driving this instability: (i) the flow-aligning parameter $\beta$, (ii) the self-propulsion strength $\alpha$, and (iii) the Frank elastic constant $K$. Self-propulsion, $\alpha {\bf p}$, aligns the velocity with the polarization, and conversely, the coupling to the flow, $\beta \bf{v}$, aligns the polarization with the velocity (see schematic in Fig.~\ref{fig: Mechanism Schematic}a).
When spatial variations in polarity occur, a competition arises between these aligning tendencies and the Frank elastic energy, which penalizes gradients in the polarization field (Fig.~\ref{fig: Mechanism Schematic}b). To illustrate this, consider an initially aligned velocity field with a polarity exhibiting a gradient along the $x$-axis (Fig.~\ref{fig: Mechanism Schematic}c). The self-propulsion and flow-alignment terms act to align velocity and polarity locally (Fig.~\ref{fig: Mechanism Schematic}d). However, this alignment frustrates the Frank elasticity, which favors a uniform polarization direction to minimize inhomogeneities. This competition creates a feedback loop: local misalignments lead to velocity changes that further distort the polarity field, amplifying fluctuations rather than suppressing them. Importantly, the instability arises because the flow alignment $\beta$ couples the polarity and velocity fields dynamically, and the absence of instantaneous force transmission allows these distortions to grow.

{\em Discussion:-}\label{summery} Our results demonstrate that a suspension of active self-propelled particles that are in contact with a substrate, the self-propulsive force alone is enough - in the absence of any dipolar active stress - to induce bend instabilities. This is noteworthy, since the hydrodynamics bend instability of a suspension of active particles is typically associated with active dipolar stresses.

The results from the numerical simulations are in agreement with the linear stability analysis, where we identify a critical self-propulsion coefficient,
$\alpha_c$ that outlines the stable and unstable states. 

We note an important distinction between our results and those reported in Ref.~\cite{MaitraPRL2020}. Maitra \emph{et al.} analyze fluctuations about a fully aligned polar state with a prescribed nonzero velocity, whereas here we perform the stability analysis about the isotropic, zero–velocity state. The latter is a natural reference configuration that does not consider perfect alignment. Furthermore, Ref.~\cite{MaitraPRL2020} eliminates velocity dynamics in the strict Stokes/adiabatic limit, while in our simulations we retain the unsteady term $\partial_t \vec{u}$ to track the finite relaxation of the flow and to decouple velocity–polarity modes. Crucially, this does \emph{not} imply an inertia–driven mechanism: in all simulations $\mathrm{Re}<10^{-3}$. Retaining the unsteady term simply allows us to follow the temporal relaxation of the flow and to decouple velocity and polarity modes. Most importantly, our fully resolved simulations demonstrate that even when initialized in the aligned moving state considered by Ref.~\cite{MaitraPRL2020}, the system destabilizes once fluctuations and nonlinearities are included.

Our findings are potentially applicable to hydrodynamic instabilities and the early stages of collective motion in systems that share the essential physical ingredients captured in our model: polar order, self-propulsion, flow alignment, substrate friction, and hydrodynamic coupling within effectively incompressible films. This description encompasses a wide range of systems such as motility assays of polar filaments~\cite{SchallerNature2010}, migratory cells in monolayers~\cite{blanch2017hydrodynamic,HeinrichElife2020,guillamat2022integer}, motile bacteria in two-dimensional setups~\cite{turiv2020polar,NishiguchiPRE2019,XuNatCom2019}, and active colloids~\cite{DeseignePRL2010, BrotoPRL2013, WeberPRL2013} that move on substrates and tend to align with the local flow direction. In such systems, flow-alignment interactions are not merely a modeling assumption but arise naturally due to the presence of a physical substrate, which mediates frictional coupling and reinforces alignment between polarity and flow~\cite{baconnier2025self}. While prior work has often interpreted observed instabilities primarily through the lens of active stress–driven mechanisms, our analysis reveals a complementary and robust route to bend instabilities emerging from the interplay of self-propulsion and flow alignment. It is important to emphasize that this mechanism is distinct from the classic active nematic instability, which arises from gradients of dipolar active stress that generate destabilizing flows. These two routes to instability, one rooted in self-propulsion and flow alignment, the other in dipolar stress and flow gradient-alignment feedback, may coexist or compete, depending on system parameters. Distinguishing between them requires careful analysis of polarity dynamics and their relation to local flow versus strain rate fields. In this context, individual-based hydrodynamic models such as squirmers~\cite{blake1971spherical}, which allow for independent tuning of self-propulsion and dipolar activity~\cite{pedley2016spherical}, offer a promising platform for systematically probing the interplay between these mechanisms. Likewise, direct measurements of active forces in experiments could help elucidate the relative contributions of polar and dipolar activity to the observed dynamics.

Finally, while the experimental confirmation of the polarity-driven instability mechanism has been limited by challenges in simultaneously quantifying polarity and flow fields, recent advances in imaging and analysis methods, specially in quantifying polarity of cells and bacteria~\cite{rodriguez2014organization,wheeler2024individual,han2025local} increasingly provide the tools to test these predictions directly.
\\

\section{Acknowledgements}
A. D. acknowledges funding from the Novo Nordisk Foundation (grant No. NNF18SA0035142 and NERD grant No. NNF21OC0068687), Villum Fonden (Grant no. 29476), and the European Union (ERC, PhysCoMeT, 101041418). Views and opinions expressed are however those of the authors only and do not necessarily reflect those of the European Union or the European Research Council. Neither the European Union nor the granting authority can be held responsible for them. The Tycho supercomputer hosted at the SCIENCE HPC center at the University of Copenhagen was used for supporting this work.

\bibliographystyle{apsrev4-1}
\bibliography{references} 

\appendix
\label{appendix}

\begin{widetext}

\section{Linear stability analysis of ordered state}
\label{stability1}

We present details of the linear stability analyses, demonstrating the robustness of the self-propulsion-induced bend instability when accounting for co-moving and co-rotational derivatives, passive elastic stresses, and also concentration variations.
The generic equation for polarization, ${\bm p}$  is, 

\begin{equation}
\partial_t {p}_i + v_k \partial_k p_i -\lambda E_{i j} p_j+\Omega_{i j}p_j=\frac{1}{\gamma}{h}_i + \beta v_i,
\label{eq: 1}
\end{equation}
where $\gamma$ is the rotational viscosity that controls relaxation of the polarity field to the minimum of the free energy through the molecular field $h_i=-\delta \mathcal{F}/\delta{p_i}$. The first two terms on the lhs of the Eq.~\ref{eq: 1} define the total derivative for the polarity field. $\lambda$ is the tumbling parameter. $\Omega_{i j}=\frac{1}{2}(\partial_i v_j - \partial_j v_i)$ and $E_{ij}=\frac{1}{2}(\partial_i v_j + \partial_j v_i)$ are the vorticity and strain rate tensors. The coefficient $\beta$ in Eq.~\ref{eq: 1}, aligns the polarization vector with the local suspension velocity \cite{MaitraPRL2020} and is specific to systems in contact with a substrate \cite{ NKumarNatCom2014, BrotoPRL2013, DadhichiJSM2018}. In this full formulation the alignment to the velocity competes with the alignment to the velocity gradients.
%
%

The equation governing the dynamics of the velocity field is given by the Navier-Stokes equation,
\begin{equation}
\rho(\partial_t + v_k\partial_k)v_i =  \partial_j \sigma_{i j} + \alpha p_i -f v_i +\beta h_i,
\label{eq: 3}
\end{equation}
where, $\eta$ is the viscosity, $f$ is the damping/friction coefficient of the substrate. $\alpha {\bm p}$ is the only active term characterizing the self propulsion in the system that describes the migration of active particles along their polarity direction ${\bm p}$.  The coefficient $\alpha$ is referred to as the self-propulsion strength. In the absence of any active force, the term $\beta {\bm h}$ is required to fulfill the detailed balance condition compensating for the $\beta {\bm v}$ term in Eq.~\ref{eq: 1} \cite{MaitraPRL2020}. Finally, $\sigma_{i j}=\sigma_{i j}^{\text{viscous}}+\sigma_{i j}^{\text{\text{elastic}}}$ is the total stress tensor, where $\sigma_{i j}^{\text{viscous}}=2 \eta E_{i j}$ is the viscous stress and $\sigma_{i j}^{\text{elastic}}=-P \delta_{i j}+\frac{1-\lambda}{2}p_i h_j - \frac{1+\lambda}{2}p_j h_i + \frac{\lambda}{2} p_k h_k \delta_{i j} $ is the passive elastic stress.\\

Here we show detailed calculations of the linear stability analysis for the full model (Eqs.~\ref{eq: 1} and ~\ref{eq: 3}). To enforce incompressibility, we introduce the stream function $\psi$, such that $v_x = \partial_y \psi$ and $v_y = -\partial_x \psi$. The vorticity is defined as a dynamical variable by $\Omega = \partial_x v_y - \partial_y v_x$. These two fields are coupled through the Poisson equation $\Delta \psi = -\Omega$, which in Fourier space yields the relation $\psi_{nm} = q^{-2} \Omega_{nm}$, where $q^2 = q_n^2 + q_m^2$. We analyze the stability of the flowing state for an infinitesimal perturbation, $\delta \phi^{(1)}(\bm r,t)=(\delta p_x, \delta p_y,\delta \Omega)=(\delta p, \theta, \Omega) $ applied on flowing state, $ \phi^0=(p_{x}^0, p_{y}^0, \Omega^0)=(1,0,0)$, such that the perturbed state is defined as $\phi(\bm r,t)=\phi^0+\delta \phi^{(1)}=(1+\delta p, \theta, \Omega)$. 
The linear order components of the molecular field  are given by, 
$$h_x  = (-8C+K \nabla^2)\delta p , \, \,\, \, \, \,h_y  = K\nabla^2 \theta .$$

Starting off we write the x-component of the polarization dynamics equation (Eq.~\ref{eq: 1}),

$$ \partial_t {p}_{x}  -  \lambda (E_{xx} p_x +  E_{xy} p_y) + \Omega_{xx} p_x + \Omega_{xy} p_y=\frac{1}{\gamma}{h}_{x}    + \beta v_x . $$

Neglecting the non-linear terms,

$$ \partial_t p_x=\frac{1}{\gamma}h_x + \beta v_x+ \lambda E_{xx}p_x $$

and introducing the perturbation one obtains,

$$ \partial_t \delta p =\frac{-8C + K \nabla^2 }{\gamma} \delta p   + \beta \partial_y \psi  +\lambda \partial_{xy} \psi.$$

To finalize, the FT in spatial coordinates is taken and the linearized equation is obtained


\begin{equation}
\partial_t \delta p ({\bf q},t) =-\frac{8C + K q^2 }{\gamma} \delta p({\bf q})   + \frac{q_m}{q^2}(\beta i - \lambda q_n) \Omega ({\bf q}).
\label{eq: apx1}
\end{equation}

Proceeding in similar fashion, the y-component of the polarization can be written as

$$ \partial_t {p}_{y}  -  \lambda (E_{yx} p_x +  E_{yy} p_y) + \Omega_{yx} p_x +\Omega_{yy} p_y=\frac{1}{\gamma}{h}_{y}    + \beta v_y $$

to the linear order,

$$ \partial_t p_y=\frac{1}{\gamma} h_y +\beta v_y + \lambda E_{yx} p_x-\Omega_{yx}p_x$$

introducing the perturbation,

$$ \partial_t \theta = \frac{K\nabla^2}{\gamma}\theta -\beta \partial_x \psi +\frac{\lambda}{2}(\partial_y^2 \psi-\partial_x^2 \psi)+\frac{\Omega}{2}.$$

Now one can take the FT on the linearized equation retrieving


\begin{equation}
\partial_t \theta ({\bm q},t)=-\frac{Kq^2}{\gamma}\theta({\bf q})- \left( -\frac{1}{2}+\frac{1}{q^2}[\beta q_n i-\frac{\lambda}{2}(q_n^2-q_m^2)]\right) \Omega({\bf q}).
\label{eq: apx2}
\end{equation}

Let us start with the velocity dynamics equation (Eq.~\ref{eq: 3}),

$$\rho(\partial_t + v_k\partial_k)v_i = \eta \nabla^2 v_i  + \partial_j \sigma^{\text{\text{elastic}}}_{ij}+\alpha p_i  -f v_i +\beta h_i .$$

Which can be written in terms of vorticity,

$$ \rho \partial_t \Omega= \partial_{xy}(\sigma_{yy}^{\text{elastic}}-\sigma_{xx}^{\text{elastic}})+\partial_x^2 \sigma_{yx}^{\text{elastic}}-\partial_y^2 \sigma_{xy}^{\text{elastic}}+\alpha (\partial_x p_y -\partial_y p_x) +\beta(\partial_x h_y -\partial_y h_x) -f \Omega +\eta \Delta \Omega. $$

To proceed it can be useful to present the linearization of the passive stress components,

$$
\sigma^{\text{elastic}}_{xx}=-\frac{1}{2}\lambda(-8C + K \nabla^2) \delta p -P,\,  \, \,\,\ \sigma^{\text{elastic}}_{yy}=\frac{1}{2}\lambda(-8C + K \nabla^2) \delta p -P, \,  \, \,\,\, \sigma^{\text{elastic}}_{xy}=\frac{1-\lambda}{2}K \nabla^2 \theta, \, \,\,\, \sigma^{\text{elastic}}_{yx}=-\frac{1+\lambda}{2}K \nabla^2 \theta.
$$

Introducing the perturbation into the vorticity equation one obtains,

\[
\begin{split}
    \rho \partial_t \Omega =\partial_{xy}(\lambda[-8C+K \nabla^2]\delta p)+\partial_x^2\left(-\frac{1+\lambda}{2}K \nabla^2 \theta \right)-\partial_y^2\left(\frac{1-\lambda}{2}K \nabla^2 \theta \right)\\
    +\alpha(\partial_x \theta-\partial_y \delta p)+ \beta[\partial_x (K \nabla^2 \theta)-\partial_y(-8C+K \nabla^2) \delta p]-f \Omega + \eta \Delta \Omega.
\end{split}
\]

Subsequently we can take the FT retrieving the following expression

\begin{equation}
\begin{split}
    \rho \partial_t \Omega ({\bm q},t)=q_m\left[ \lambda(8C +Kq^2) q_n -i (\alpha - \beta(8C+Kq^2)) \right]\delta p({\bm q}) \\
    +\left[\frac{Kq^2}{2}(\lambda(q_m^2-q_n^2)-q^2)+i q_n (\alpha -\beta K q^2)\right] \theta({\bm q}) -(\eta q^2 +f) \Omega({\bm q})   
\end{split}
\label{eq: apx3}
\end{equation}





From the linearized equations (Eqs. \ref{eq: apx1}, \ref{eq: apx2} and \ref{eq: apx3}) we obtain the Jacobian matrix:

\begin{equation}
\scriptsize
\centerline
{${\mathcal{ J }}(q_n,q_m) = \begin{bmatrix}
    -\frac{8C+Kq^2}{\gamma}& 0 & \frac{q_m}{q^2}(i \beta-\lambda q_n) \\
    0 & -\frac{K q^2}{\gamma} & \frac{1}{2}-\frac{1}{q^2}(\beta q_n i -\frac{\lambda}{2}(q_n^2-q_m^2))  \\
    \frac{q_m}{\rho} \left[\lambda (8C+Kq^2)q_n -i(\alpha- \beta(8C+Kq^2))\right] &\frac{1}{\rho}[\frac{Kq^2}{2}(\lambda(q_m^2-q_n^2)-q^2)+i q_n(\alpha -\beta K q^2)] & -\frac{1}{\rho}(\eta q^2+f)  
\end{bmatrix}$}
\label{eq: apx5}
\end{equation}

For the longitudinal mode $(n,m)=(1,0)$ or $(q_n,q_m)=(q,0)$, the coupling is given by $(\theta,\Omega)$ and the matrix simplifies to,
\begin{equation}
    \label{eq: longitudinal mode}
     \mathcal{J}_{10}(q,0)=
    \begin{bmatrix}
        -\frac{K q^2}{\gamma} &\frac{1+\lambda}{2} -i \frac{\beta}{q}\\
        -\frac{Kq^4}{2 \rho}(1+\lambda)+i \frac{q}{\rho}(\alpha -\beta Kq^2)  & -\frac{1}{\rho}(\eta q^2+f)
    \end{bmatrix}.
\end{equation}
For the transverse mode, $(n,m)=(0,1)$ or $(q_n,q_m)=(0,q)$ the Jacobian matrix simplifies to,
\begin{equation}
    \mathcal{J}_{01}(0,q)=
    \begin{bmatrix}
        -\frac{8C +K q^2}{\gamma} & 0 & i \frac{\beta}{q}  \\
        0 & -\frac{K q^2}{\gamma} & -\frac{\lambda-1}{2} \\
        i \frac{q}{\rho}[\alpha -\beta(8C +Kq^2)] & \frac{K q^4}{2 \rho}(\lambda-1) & -\frac{1}{\rho}(\eta q^2+f)
    \end{bmatrix}
\end{equation}
which couples $(\delta p, \theta, \Omega)$. The longitudinal mode resembles the result obtained in the main text. By neglecting higher-order terms in each matrix component, we recover the same critical self-propulsion and dispersion relation as reported in the main text (Eqs.~8 and 9, main text). This proves that the instability persists unchanged in the full model, when the co-moving, co-rotational derivatives, and elastic stresses are included.

\section{Concentration effect on the linear stability}
\label{ConcEffect}
\ndg{In this section, we present a detailed linear stability analysis of the system when polarity and flow are coupled to the particle concentration field, $c$, showing that introducing concentration as a dynamic variable does not suppress the instability examined in this manuscript. To display this, we introduce an advective diffusive equation of the particle concentration coupled with the constitutive equations (Eq.~\ref{eq: 1} and Eq.~\ref{eq: 3}),}

\begin{equation}
    \centering
    (\partial_t  + v_k \partial_k) c= - \partial_k (w_2 c p_k - D \partial_k c).
\label{eq: 101}    
\end{equation}

\ndg{Moreover, we incorporate a term in the free energy that accounts for the coupling between the concentration and the splay deformation, $\mathcal{F}_c =-\int dA  [w_3 \frac{c-c_0}{c_0} \nabla \cdot p]$. Introducing concentration as a new dynamic variable and accounting for perturbations with respect to the mean concentration, $c= c_0 + \delta c$, one can find the new Jacobian matrix,}

\begin{equation}
    \label{eq: 102}
    \scriptsize
\centerline
    {${\mathcal{ J }}(q_n,q_m) = \begin{bmatrix}
    -D q^2-i w_2 q_n & -i w_2 q_n c_0 & -i w_2 q_m c_0 & 0 \\
    -\frac{i w_3 q_n}{\gamma c_0}&-\frac{8C+Kq^2}{\gamma}& 0 & \frac{q_m}{q^2}(i \beta-\lambda q_n) \\
    -\frac{i w_3 q_m}{\gamma c_0}&0 & -\frac{K q^2}{\gamma} & \frac{1}{2}-\frac{1}{q^2}(\beta q_n i -\frac{\lambda}{2}(q_n^2-q_m^2))  \\
    \frac{\lambda-1}{2\rho}\frac{w_3}{c_0}i q_m q
    ^2 & \frac{q_m}{\rho} \left[\lambda (8C+Kq^2)q_n -i(\alpha- \beta(8C+Kq^2))\right] &\frac{1}{\rho}[\frac{Kq^2}{2}(\lambda(q_m^2-q_n^2)-q^2)+i q_n(\alpha -\beta K q^2)] & -\frac{1}{\rho}(\eta q^2+f)  \end{bmatrix}.$}
\end{equation}

\ndg{Notably, for the longitudinal mode, $(n,m)=(1,0)$ the Jacobian matrix converts to a block diagonal matrix,}
\begin{equation}
    \mathcal{J}(q,0)=\begin{bmatrix} A & 0 \\
    0 & B
    \end{bmatrix}
\end{equation}
where $A$ takes the following form
\begin{equation}
    A=\begin{bmatrix}
        -Dq^2-i w_2 q & -i w_2 q c_0 \\
        -\frac{i w_3 q}{\gamma c_0} & -\frac{8C + K q^2}{\gamma}
    \end{bmatrix},
\end{equation}
and $B$ takes
\begin{equation}
    B=\begin{bmatrix}
         -\frac{K q^2}{\gamma} &\frac{1+\lambda}{2} -i \frac{\beta}{q}\\
        -\frac{Kq^4}{2 \rho}(1+\lambda)+i \frac{q}{\rho}(\alpha -\beta Kq^2)  & -\frac{1}{\rho}(\eta q^2+f)
    \end{bmatrix}.
\end{equation}
\begin{figure}[h!]
\centering
\includegraphics[width=0.75\linewidth]{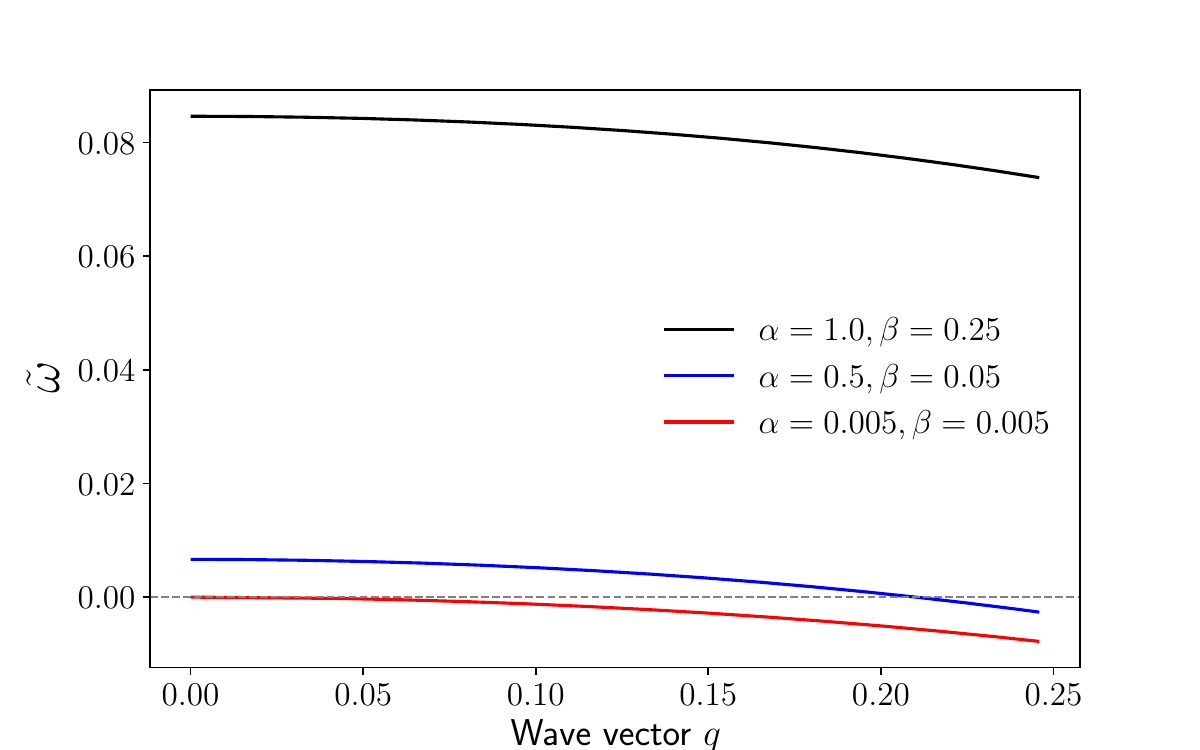}
\caption{\textbf{Growth rate curves of the system coupled to concentration dynamics.} Growth rate, $\tilde{\omega}(q)$ versus wave-vector $q$ plotted for the longitudinal mode (Eq.~\ref{eq: 102}) for different $\alpha$ and $\beta$ values.}
\label{fig: supfig102}
\end{figure}

\ndg{We note that the eigenvalues of the longitudinal modes are decoupled: two of them involve the coupling between the x-component of polarity and the concentration field ($p_x$ and $c$), while the other two involve the coupling between the y-component of polarity and the vorticity ($p_y$ and $\Omega$). Matrix $B$ is identical to the Jacobian obtained from the full model (Eq.~\ref{eq: longitudinal mode}); therefore, the stability diagram and growth rate of the longitudinal mode remain unchanged compared to the model without concentration. This implies that concentration does not alter the instability.} \ndg{To further demonstrate this we show the growth rate, $\tilde{\omega}$ obtained from the Jacobian matrix (Eq.~\ref{eq: 102}) for three different sets of $(\alpha, \beta)$. For large values of self-propulsion and flow coupling coefficient the growth rate remains positive, $\tilde{\omega} >0$ (\textit{blue and black line }Fig.~\ref{fig: supfig102}). However, for smaller values of these parameters the growth rate remains negative for all wave-vectors, $\tilde{\omega} <0$ (\textit{red line} Fig.~\ref{fig: supfig102}). We have also confirmed through full numerical simulations of the active polar system coupled to concentration variations that the bend stability occurs even when the dynamics of the concentration field is included, in agreement with the linear stability analyses.}

\section{Analytical growth rate relation of the main text}
\label{FullOmega}
\ndg{In this section we present the analytical expression of the growth rate that is discussed in the main text. The growth rate is obtained by solving the second-order algebraic equation in Eq.~8 of the main text, yielding the following expression: }

\begin{equation}
    \label{eq: analytical expression for the growth rate}
    \ndg{\tilde{\omega}_{\pm} = \frac{-\left(f + \rho \frac{K}{\gamma} q^2\right) \pm \sqrt{\left(f + \rho \frac{K}{\gamma} q^2\right)^2 - 4 \rho \left[Kq^2 \left( \frac{f}{\gamma}+\beta^2\right) - \beta \alpha\right]}}{2 \rho}.}
\end{equation}

\ndg{In the main text, we focus on $\tilde{\omega}_+$ since it corresponds to the growth rate that can become positive.}
\end{widetext}

\end{document}